\documentclass[letterpaper,twocolumn,english,aps,prl]{revtex4}
\usepackage{ae,aecompl}
\usepackage[T1]{fontenc}
\usepackage{color}
\usepackage{textcomp}
\usepackage{amssymb}
\usepackage{graphicx}

\makeatletter

\pdfpageheight\paperheight
\pdfpagewidth\paperwidth

\DeclareRobustCommand{\greektext}{%
  \fontencoding{LGR}\selectfont\def\encodingdefault{LGR}}
\DeclareRobustCommand{\textgreek}[1]{\leavevmode{\greektext #1}}
\DeclareFontEncoding{LGR}{}{}
\DeclareTextSymbol{\~}{LGR}{126}

\@ifundefined{textcolor}{}
{%
 \definecolor{BLACK}{gray}{0}
 \definecolor{WHITE}{gray}{1}
 \definecolor{RED}{rgb}{1,0,0}
 \definecolor{GREEN}{rgb}{0,1,0}
 \definecolor{BLUE}{rgb}{0,0,1}
 \definecolor{CYAN}{cmyk}{1,0,0,0}
 \definecolor{MAGENTA}{cmyk}{0,1,0,0}
 \definecolor{YELLOW}{cmyk}{0,0,1,0}
 }

\makeatother

\usepackage{babel}
\begin{document}

\title{Monoatomic magnetic interfaces in IrMn/Cr/Co thin films probed by
grazing incidence X-ray absorption spectroscopy}

\author{I. García-Aguilar}

\author{N. M. Souza-Neto}

\affiliation{Laboratório Nacional de Luz Síncrotron (LNLS), Campinas, 13084-971
São Paulo, Brazil}

\author{G. M. Azevedo}

\author{S. Nicolodi}

\author{L. G. Pereira}

\author{J. E. Schmidt}

\author{J. Geshev}

\affiliation{Instituto de Física, UFRGS, Porto Alegre, 91501-970 Rio Grande do
Sul, Brazil}

\author{C. Deranlot}

\author{F. Petroff}

\affiliation{Unité Mixte de Physique CNRS/Thales, 91767 Palaiseau and Université
Paris-Sud, 91405 Orsay, France}

\date{\today{}}

\begin{abstract}
We present depth-resolved experimental results on the atomic and electronic
structures of the Co-Cr interface on four IrMn/Cr/Co thin films with
variable thickness of the Cr layer. Grazing incidence X-ray absorption
near edge structure near the Cr K-edge was used, and an Ångstrom resolved
depth-profile for this layer was obtained. An interdiffusion between
chromium and cobalt layers was observed in all films, being more pronounced
for samples with thinner Cr layers , where Cr behaves as an amorphous
material. This causes a contraction in coordination distances in Cr
near the interface with Co. In this region, a change in the electronic
structure of chromium's 3$d$ orbitals is also observed, and it appears
that Cr and Co form a covalent bond resulting in a CrCo alloy. \emph{Ab
initio} numerical simulations support such an interpretation of the
obtained experimental results. 
\end{abstract}
\maketitle
\textcolor{black}{Magnetic thin-film heterostructures have been the
most common type of systems used }in the exchange bias (EB) phenomenon
research. Although the exchange coupling between the ferromagnetic
(FM) and antiferromagnetic (AF) materials has been usually considered
to be a nearest-neighbor interaction \cite{Thomas-JAP-2000,Mewes-JAP-2000,Gruyters,Wang,Yoo,Geshev-PRB07}, long
range interactions across a thin spacer layer (SL) in the FM/AF interfaces
have also been reported \cite{Gokemeijer-PRL97} . Theoretical models
\cite{Blundell-2001} indicate that the electronic structure is rather
important for RKKY-like long-range interactions , which has been supported
by experimental results on the dependence of the FM/AF exchange coupling
on the spacer layer \cite{Gokemeijer-PRL97,Nicolodi-JMMM07}. Previous
studies on IrMn/Cr/Co thin films, in particular, have shown that magnetic
phenomena such as EB and rotatable anisotropy may strongly depend
on the thickness of the Cr (a weak AF material) spacer layer thickness
as well \cite{Nicolodi-PRB2012}. In the regard of the latter, an
important part of magnetic systems research is the detailed characterization
of the electronic and atomic structures as a function of the depth
on such samples. 

Given the layered nature of these samples and the single-element spacer
layer, X-ray Absorption Spectroscopy (XAS) proves to be an effective
technique due to it's chemical selectivity \cite{Bianconi-Koningsberger88}.
Using direct incidence XAS it is not possible to probe nanometric
penetration depth; instead, it is convenient to use a grazing incidence (GI)
geometry since it allows a selective peer into the depth of the sample
\cite{Souza-Neto-JAC09,Souza-Neto-APL06}. For thin film studies,
this confinement also has the considerable advantage of minimizing
other layers' contributions as well. Near edge structure on XAS (XANES)
provides extra sensitivity information on the closest-surrounding
atomic arrangement and on density of unoccupied states, i.e., on the
electronic structure. Using XANES combined with the GI geometry, it
is possible to obtain detailed information for the atomic, electronic
and magnetic structural differences of monoatomic sublayers across
the chromium SL. 

Here we present a qualitative depth profile of the atomic and electronic
structure of the chromium layers in four polycrystalline IrMn/Cr($\mathrm{t_{Cr}}$)/Co
thin films, where the Cr thickness $\mathrm{t_{Cr}}$ is varied. A
systematic scan of these layers was done using GI-XAS spectroscopy
in the XANES regime. Based on collected spectra near the chromium
edge, the structural information obtained for the Co/Cr interfaces
is analyzed. In addition to the experimental results, interface properties
were further examined using \emph{ab initio} simulations.

The four thin films with the composition Ru(150 Å)/IrMn(150 Å)/Cr($\mathrm{t_{Cr}}$)/Co(50
Å)/Au(100 Å) were deposited by magnetron sputtering onto Si(100) substrates,
where $\mathrm{t{}_{Cr}}$=2.5, 7.5, 10 and 20 Å. Complete description
of the samples' preparation can be found in Ref. \cite{Geshev-PRB07}.
Since the Cr thickness is significatively low for all films, Ångstrom
resolution is needed to accurately study the whole chromium layer.
We employed GI-XANES for obtaining this resolution using the method
described in Ref. \cite{Souza-Neto-JAC09}. 

\begin{figure}
\includegraphics[scale=0.3]{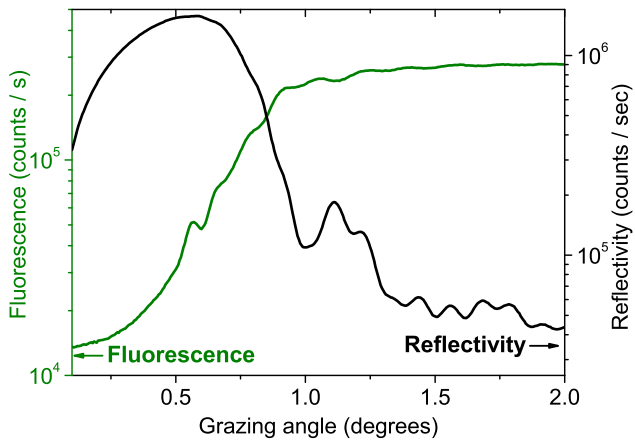}\caption{\label{fig-reflet-fluor} (Color online) Fluorescence and reflectivity
data as a function of grazing angle $\theta$, at a fixed energy
above the Cr absorption threshold, with approximate maximum absorbance. }
\end{figure}

The GI-XANES measurements were performed at the Brazilian Synchrotron
Light Laboratory (LNLS), using the XAFS2 beamline. XANES spectra were
collected and normalized for all thin films near the Cr K-edge (\emph{E}
= 5989 eV). Energy calibration was done using a reference chromium
foil. The incident beam with electric field along the thin-film surface
plane was focused on the sample with a vertical divergence of about
0.01$^{\circ}$, which intensity was monitored using a first ion chamber.
The reflected beam and the fluorescence emission were simultaneously
collected using a second ion chamber and a 15-element Ge detector,
respectively. The working grazing angle calibration was done for every
sample using an angular profile at a fixed energy like that shown
in Fig.\ref{fig-reflet-fluor}. The angle values were chosen around
the fluorescence decreasing region. 

\begin{figure}
\includegraphics[scale=0.3]{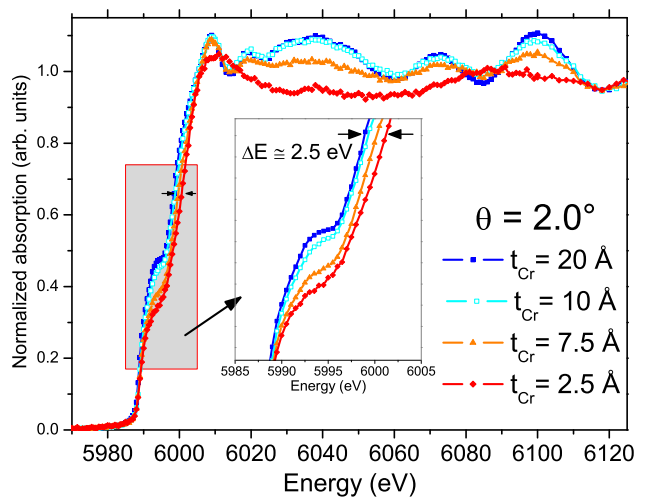}\caption{\label{fig-xanes-samples} (Color online) Comparison of XANES spectra
measured in samples with different Cr layer's thickness in the highest
measured angle $\theta$ = 2.0$^{\circ}$. }
\end{figure}

The fluorescence curves used for calibration indicated a complete
scan of the chromium layer for all samples when $\theta$ = 2.0$^{\circ}$.
XANES spectra for this grazing angle for each sample are shown in
Fig. \ref{fig-xanes-samples}. The low absorption edge from Cr allows
for XANES contributions from this layer only, even at high angles.
Thicker layers (symbols\textcolor{blue}{{} $\blacksquare$} and\textcolor{cyan}{{}
$\square$}) have a metallic chromium structure which is remarkably
different from that of the films with $\mathrm{t_{Cr}}$ = 2.5 and
7.5 Å, where the metallic configuration is lost and more amorphous-like
spectra are observed. Since high angle measurements show a convolution
of all layers' contributions, in these thinner Cr layers' samples
practically the whole Cr layer is interdiffused into the other layers.

The lower inset in Fig. \ref{fig-xanes-samples} shows a substantial
energy shift (\textgreek{D}\emph{E}\ensuremath{\approx}2.5 eV), meaning
that near the Cr/Co interface more energy is needed to overcome near
atomic potentials \cite{Souza-Neto-PRB2004} which could be interpreted
as a contraction in the coordination distances in this top sublayer,
and thus an interdiffusion between the chromium and cobalt layers.
For samples with thicker Cr layer this interdiffusion is also clearly
observed at the Co/Cr interface as will be discussed below. 

\begin{figure*}
\includegraphics[scale=0.21]{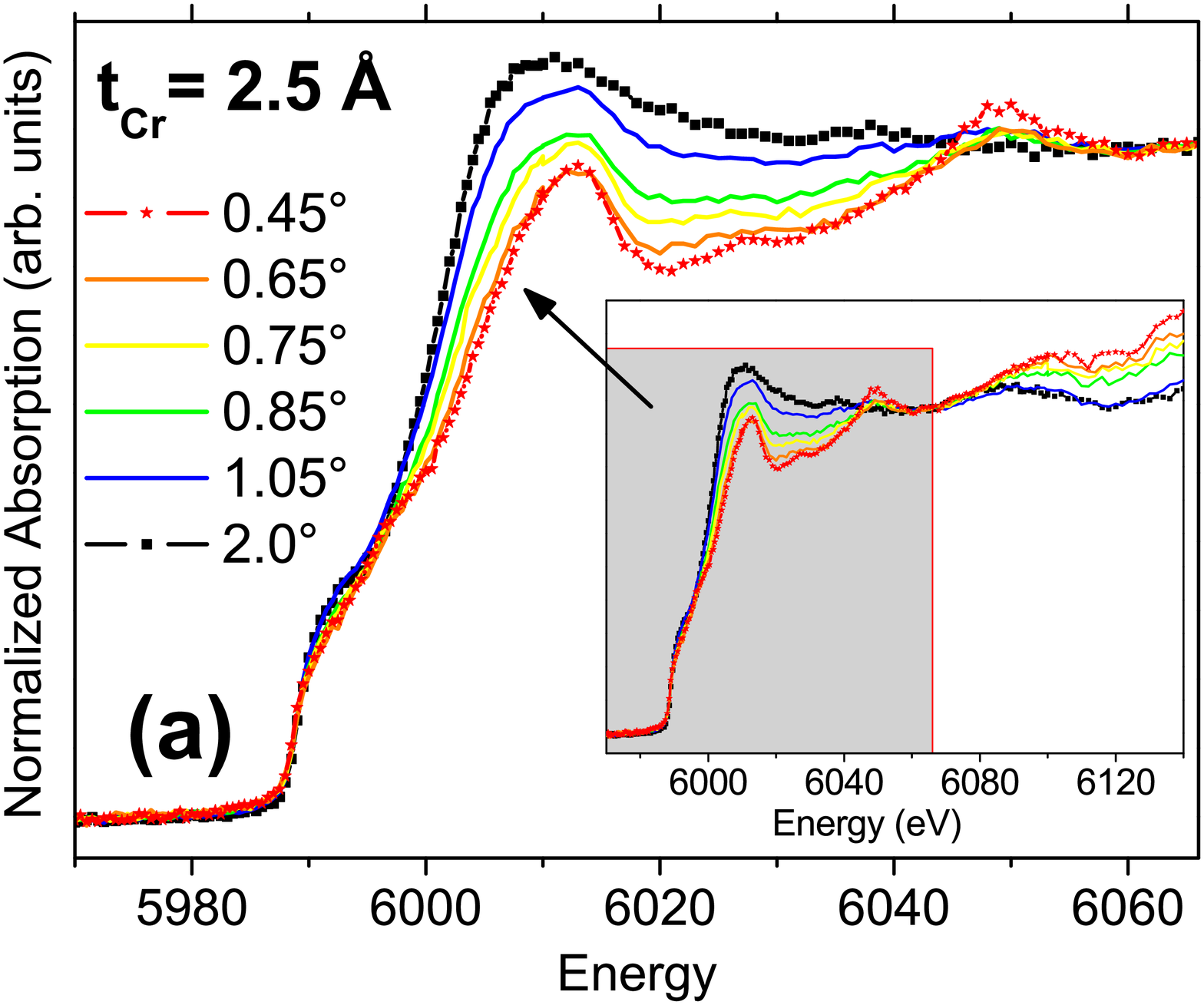}\includegraphics[scale=0.21]{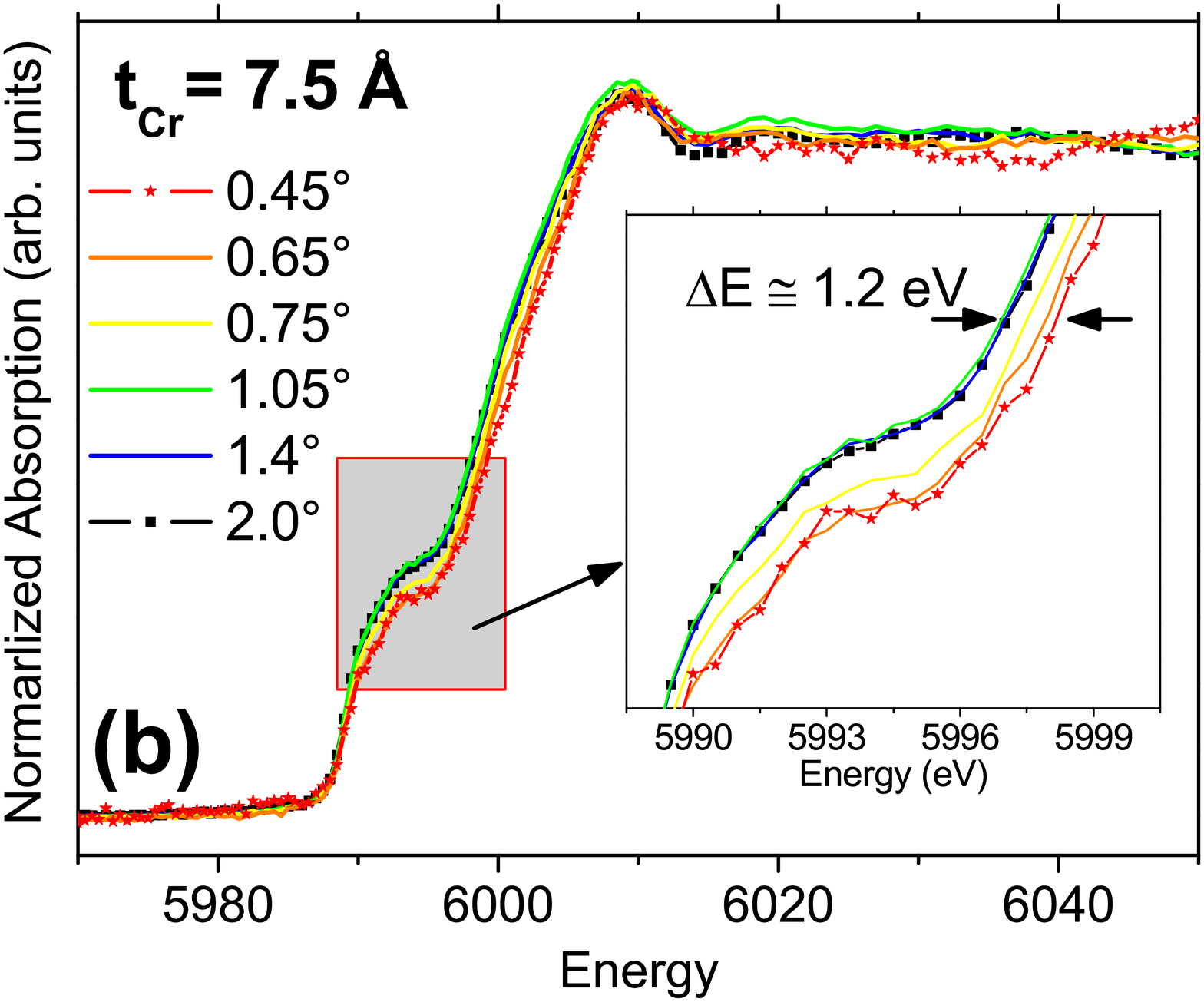}\includegraphics[scale=0.21]{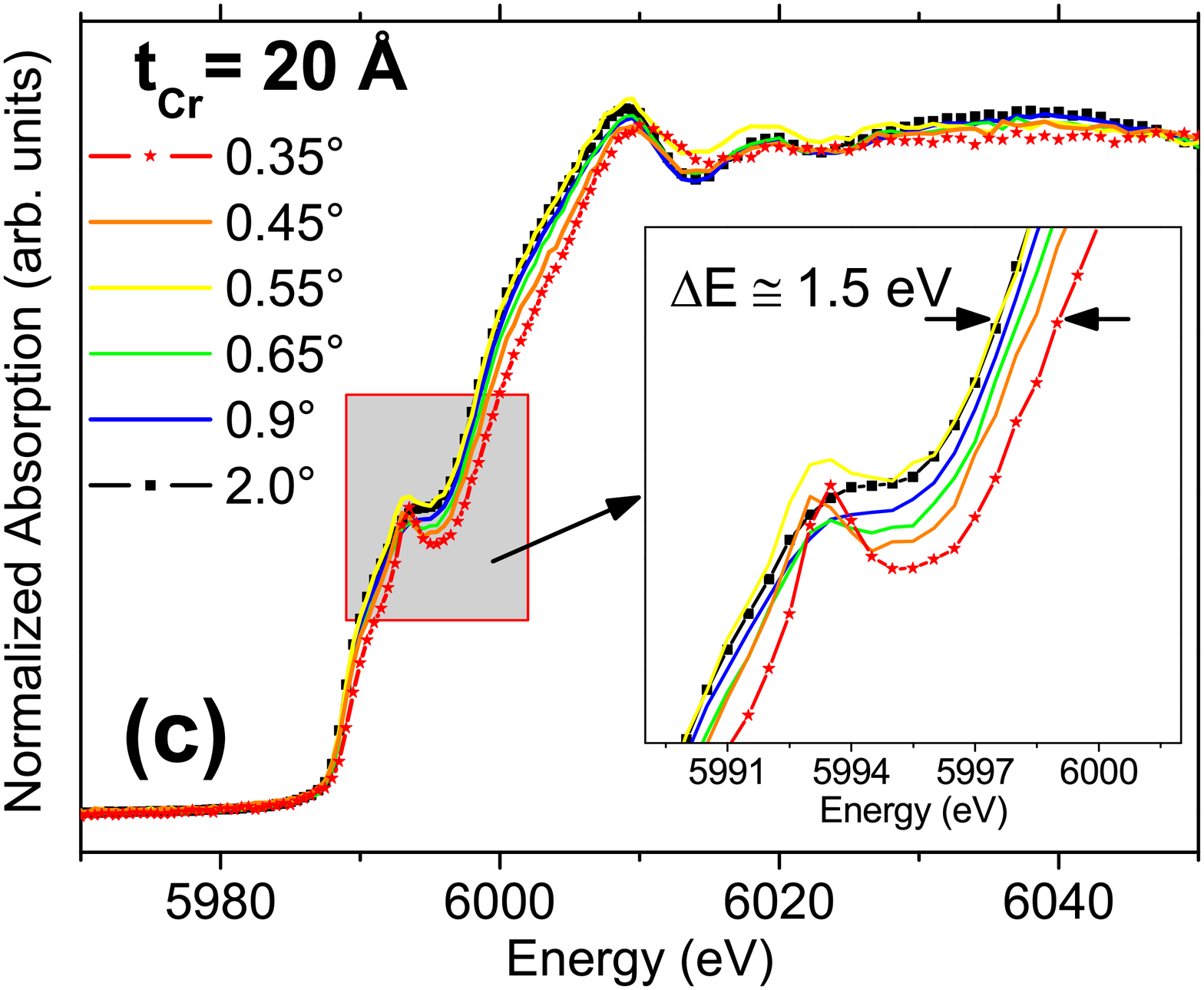}\caption{\label{fig-xanes-each-sample} (Color online) XANES spectra as a function
of the grazing angle for samples with different Cr layer thicknesses:
(a) 2.5 Å, (b) 7.5 Å and (c) 20 Å. }
\end{figure*}

Besides this change in the absorption edge, there is a progressive
decrease in the pre-edge intensity as the chromium layer becomes thinner,
and thus as contributions from Cr/Co interface are most appreciable.
Given the dependence of the absorption coefficient $\mu$ on the density
of unoccupied states \cite{Bianconi-Koningsberger88}, a drop of the
absorption intensity means an increase in occupied states. In XANES,
the pre-edge region has a strong contribution from the quadrupole
terms (1\emph{s} -> 3\emph{d}, in this case), and thus, it probes
empty electronic states in 3\emph{d} orbital. Because its intensity
decreases in the Cr/Co interface, it appears that chromium (Cr:{[}Ar{]}$4s^{1}3d^{5}$)
forms a covalent bond with cobalt (Co:{[}Ar{]} $4s^{2}3d^{7}$), resulting
in an increase in occupied 3\emph{d} states in Cr.

In Fig. \ref{fig-xanes-each-sample}, the XANES spectra show a qualitative
depth profile for $\mathrm{t_{Cr}}$= 2.5, 7.5 and 20 Å, summarizing
differences in electronic and atomic structures in the chromium layer,
as a function of its thickness. Similarly to the thickness dependence
discussed above, a correlated behavior is found for each sample. Although
in Fig. \ref{fig-xanes-samples} metallic spectra were observed for
high angles for $\mathrm{t_{Cr}}$ = 10 and 20 Å, the $\theta=0.35^{\circ}$
spectrum (\textcolor{red}{$\star$}) in Fig. \ref{fig-xanes-each-sample}(c)
shows the amorphous-like structure and energy shift shows contraction
of atomic distances in all plots, meaning that interdiffusion of the
chromium and cobalt layers is present in all samples. In the sample
with $\mathrm{t_{Cr}}$ = 10 Å , the Cr layer is already thick enough
to have metallic chromium at the bottom, as evidenced by results equivalent to
those shown in Fig. \ref{fig-xanes-each-sample}(c). For these thicker
Cr layer films, the Co-Cr sublayers are considerably thinner as compared
to the metallic chromium, thus the contribution from the CrCo alloy
is negligible when scanning the whole layer. 

No crystalline structure is observed throughout the Cr layer as seen
in Fig. \ref{fig-xanes-each-sample}(a), which shows a complete interdiffusion
somewhat expected from the fact that the Cr layer thickness is extremely
small as compared to one or two monoatomic layers and that the samples
were grown by sputtering. This amorphous behavior is still observed
in Fig. \ref{fig-xanes-each-sample}(b) $\theta$<1.05$^{\circ}$ \textcolor{black}{where
EXAFS oscillations are lost in the spectra}\textcolor{green}{{} }meaning
that the layer interdiffusion has a depth of about 5 Å. This highlights
the resolution strength of the GI-XAS technique in studying thin films
and its proficiency in probing monoatomic layers. 

The inset in Fig. \ref{fig-xanes-each-sample}(a) shows a GI geometry
effect in XAS. Since the amount of the probed depth is governed by
the exponential decay of the incident radiation, the whole 2.5 Å thick
Cr layer is probed at any angle/energy but with different weighted
contributions. In other words, for low angles, when only the upper
region of the Cr layer is being probed, higher energy photons get
to penetrate deeper in the layer, which results is a progressive increase
in the absorption intensity when increasing the photon energy (as
seen in the $\theta=0.45^{\circ}$ spectrum (\textcolor{red}{$\star$})
for example). At higher angles, this intensity sensitivity on energy
becomes insignificant, since X-rays penetrate the whole sample merely
because of the geometry itself; a complete scan of the layer is being
done in all energies, and the XANES spectra do not show this energy-dependence
effect. 

\begin{figure}
\includegraphics[scale=0.15]{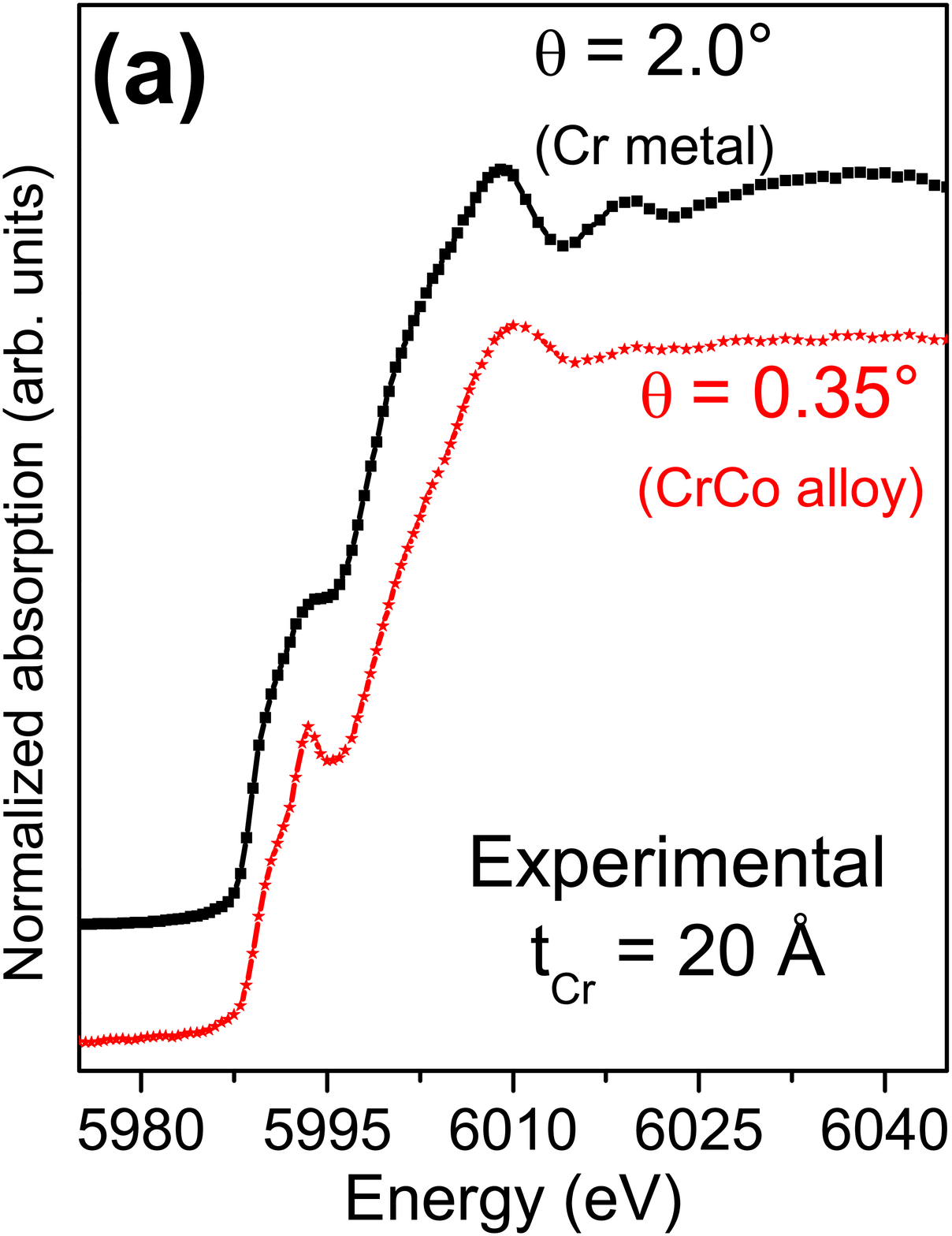}\includegraphics[scale=0.15]{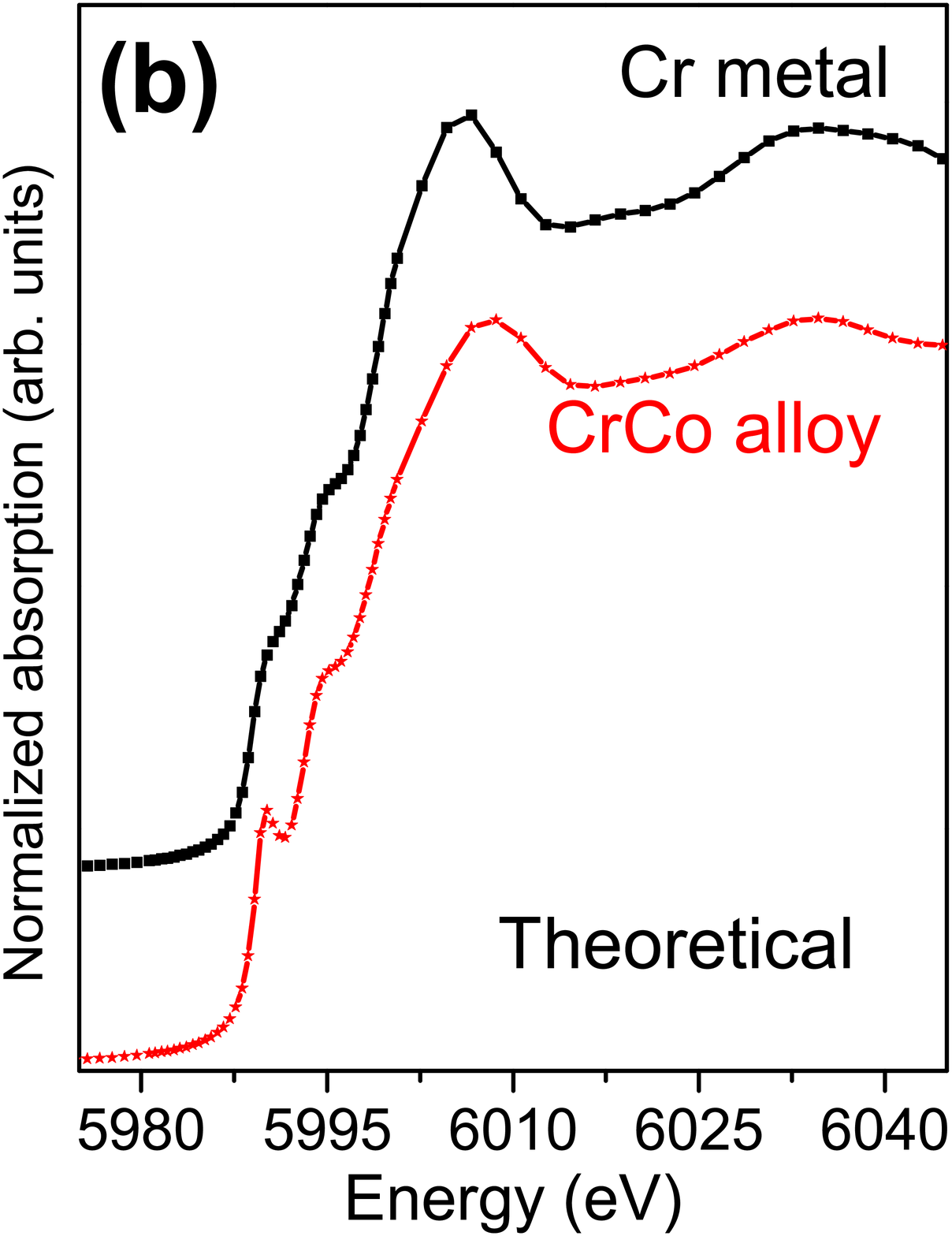}\caption{\label{fig-xanes-exp-theo} (Color online) (a) Experimental XANES
spectra for the sample with $\mathrm{t_{Cr}=20\mathring{A}}$ for
two grazing angles showing the Co/Cr interface ($\theta$ = 0.35$^{\circ}$)
and a bulk-like metallic Cr character ($\theta$ = 2.0$^{\circ}$).
(b) Numerically-simulated spectra for a metallic Cr bcc lattice and
a CrCo alloy with bcc structure with a more contracted lattice. }
\end{figure}

The differences in structure between the upper, interdiffused CrCo
layer, and the deeper metallic sublayers are clearer in Fig. \ref{fig-xanes-each-sample}(c),
especially at the pre-edge. These variations are better appreciated
in Fig. \ref{fig-xanes-exp-theo}(a), that shows the experimental
spectra measured in the lowest and highest angles only; thus, the
interface structure as interpreted above is a CrCo alloy and metallic
chromium. To support this interpretation \emph{ab initio} simulations
using FDMNES code \cite{Joly-PRB01} were performed in order to account
for a bcc Cr metal structure and a bcc CoCr alloy with smaller interatomic
distance, based on the interdiffusion contraction of chromium. Numerical
results are shown in Fig. \ref{fig-xanes-exp-theo}(b) for metallic
Cr and a CrCo alloy. In both experimental and simulated data, there
it is a smoothening of the amorphous spectra (\textcolor{red}{$\star$})
past the absorption edge when comparing them with the metallic ones
($\blacksquare$). A consequential similarity between them is the
appearance of a peak in the pre-edge region due to 3\emph{d} orbital's
changes in agreement with different electronic structures in Cr at
the Co-Cr interface resulting from a covalent bond with Co. Besides
this, the energy shift associated to the lattice contraction and present
in all samples is also found theoretically in the calculated spectra.
As a whole, the theoretical simulations agree well with the interpretation
of experimental results. 

In summary, GI-XANES near chromium K-edge was used to develop Ångstrom
resolved depth profiles of electronic and atomic structures on magnetic
Co/Cr ($\mathrm{t_{Cr}}$) /IrMn films near the Co-Cr interface, as
a function of the chromium layer thickness. Interdiffusion between
the chromium and cobalt layers was observed for all samples, resulting
in a loss of the crystalline structure near the interface. In the
samples with thinner Cr layer, i.e. $\mathrm{t_{Cr}}$ = 2.5 and 7.5
Å), the whole Cr layers lose the bcc structure behaving as amorphous
material, whereas for bigger $\mathrm{t_{Cr}}$ low angle spectra
evince an interdiffusion of 5 Å in depth. This causes a contraction
in the coordination distances near the Co-Cr interface. In this region,
there seems that Cr forms a covalent bond with Co as suggested by
a change in the electronic structure of chromium 3\emph{d} orbitals
in the Cr XANES pre-edge, which results in a CrCo alloy. Numerical
simulations for a bcc CrCo structure showed consistency with the measured
spectra. XANES spectroscopy in the grazing incidence geometry proved
to be an efficient technique for an accurate study of these magnetic
thin film.
\begin{acknowledgments}
We thank the financial support from the Brazilian agencies CNPq, CAPES
and FAPERGS.
\end{acknowledgments}
\bibliographystyle{apsrev}

\end{document}